\documentclass[graybox]{svmult}

\usepackage{type1cm}        
\usepackage{makeidx}         
\usepackage{graphicx}        
\usepackage{multicol}        
\usepackage[bottom]{footmisc}

\usepackage{newtxtext}       %
\usepackage{newtxmath}       

\makeindex             

\usepackage{caption,subcaption}

\begin{document}

\title*{The Voronoi tessellation method in astronomy}
\author{Iryna Vavilova, Andrii Elyiv, Daria Dobrycheva, Olga Melnyk}
\institute{Iryna Vavilova \at Main Astronomical Observatory of the NAS of Ukraine, 27 Akademik Zabolotny St., Kyiv, 03143, Ukraine,
\email{irivav@mao.kiev.ua}
\and Andrii Elyiv \at Main Astronomical Observatory of the NAS of Ukraine, 27 Akademik Zabolotny St., Kyiv, 03143, Ukraine, 
\email{elyiv@mao.kiev.ua}
\and Daria Dobrycheva \at Main Astronomical Observatory of the NAS of Ukraine, 27 Akademik Zabolotny St., Kyiv, 03143, Ukraine, 
\email{daria@mao.kiev.ua}
\and Olga Melnyk \at Main Astronomical Observatory of the NAS of Ukraine, 27 Akademik Zabolotny St., Kyiv, 03143, Ukraine, 
\email{melnykol@gmail.com}}
%
%
\maketitle


\abstract{The Voronoi tessellation is a natural way of space segmentation, which has many applications in various fields of science and technology, as well as in social sciences and visual art. The varieties of the Voronoi tessellation methods are commonly used in computational fluid dynamics, computational geometry, geolocation and logistics, game dev programming, cartography, engineering, liquid crystal electronic technology, machine learning, etc. 
The very innovative results were obtained in astronomy, namely for a large-scale galaxy distribution and cosmic web pattern, for revealing the quasi-periodicity in a pencil-beam survey, for a description of constraints on the isotropic cosmic microwave background and the explosion scenario likely supernova events, for image processing, adaptive smoothing, segmentation, for signal-to-noise ratio balancing, for spectrography data analysis as well as in the moving-mesh cosmology simulation. We briefly describe these results, paying more attention to the practical application of the Voronoi tessellation related to the spatial large-scale galaxy distribution.}

\section{The Voronoi tessellation in a spatial galaxy distribution: first works and basic approach}
\label{sec:1}

The geometrical methods based on the Voronoi diagram deal with a partitioning of space into regions in a specific subset of generators. It was named after Georgy F. Voronoi (April 28, 1868, Zhuravka village, Chernihiv region, Ukraine – Nov 20, 1908, Warsaw, Poland), the outstanding Ukrainian mathematician ~\cite{Syta2009, Pratsyovity2018}, who studied the general n-dimensional case of these diagrams ~\cite{Voronoi1907, Voronoi1908}. 

In 1984, Matsuda and Shima advanced the idea to apply the Voronoi tessellation method for describing the cellular structure of the local Universe ~\cite{Matsuda1984}, finding a topological tendency of galaxies ``to cluster at the vertices, edges and faces of polyhedral shaped voids''. In 1987, Ling demonstrated that the Voronoi tessellation and the Minimal Spanning Tree being applied to the CfA Redshift Survey of galaxies (the first survey to map the large-scale structure of the Universe) are able to detach filamentary structures and voids \cite{Ling1987}. In 1989, Yoshioka \& Ikeuchi proposed three-dimensional Voronoi tessellation as a model of the evolution of the negative density perturbations regions, which resulted in the overlapping of shells while the modeled skeleton can be compared with real observed structures and with mass distribution correlation functions \cite{Yoshioka1989}. 

\begin{figure}[h]
    \centering
    \includegraphics[width=1.0\textwidth]{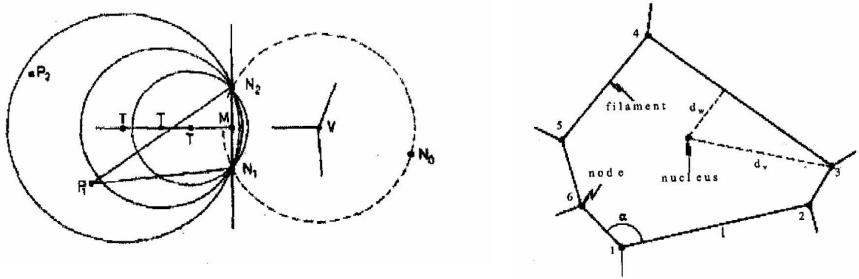}
    \caption{(Left) The construction of a new Delaunay triangle from two known nuclei $N_{3}$ such that ($N_{1}$, $N_{2}$, $N_{3}$) forms a triangle whose circumsphere does not contain any other nucleus in the Voronoi tessellation. (Right) Identification of the four quantities which were calculated in each Voronoi cell: $l_{i}$, the length of wall \textit{i}; $\alpha$, the angle between two walls meeting at vertex; $d_{w}$, the distance between the nucleus and a wall, where the projection of the nucleus doesn't necessarily lie on the wall (Icke, 1987, open astronomy).}
    \label{voronoi-pattern}
\end{figure}

For the first time, the Voronoi tessellation was considered in detail as a pattern of matter distribution in the Universe in work by Icke and Weygaert ~\cite{Icke1987} and series of their following works ~\cite{Icke1988, Weygaert1989, Icke1991}. These authors concluded that the regions of lower density become more spherical with evolution and matter floods away from expansion centers and accrues at the borders of packing of spheres. This leads to the partition of space on the Voronoi tessellation with nuclei in the centers of low-density regions called the voids. High-density regions - clusters of galaxies - lie at the crossing of vertexes of adjacent cells, filaments at the edge of cells, and pancakes of large-scale structure (LSS) are faces of cells (Fig.~\ref{voronoi-pattern}, right). Sheth et al. ~\cite{Sheth2004} have developed its idea and considered the model of a void created in the frame of the Voronoi tessellation paradigm.

The Voronoi tessellation can be constructed as follows. Let us consider a Voronoi cell of finite size in N-dimensional space (usually N = 2 or N = 3), where a fixed number of points is distributed according to some statistical law (for example, the Poisson law). Suppose that each point is the center of a spherical expanding bubble structure. If all these structures begin to expand at the same moment with the same rate, the bubbles will be touched in planes that perpendicularly bisect the lines connecting the centers of expansion. These bisecting planes, in turn, intersect each other. As a result of this process, new lines will be generated, which in turn intersect each other and form a network. Using an adopted terminology, we will call such a center of the cell as a nucleus. So, each nucleus will be enclosed by a set of (N - 1) - dimensional planes forming a convex cell. Distribution of nuclei forms the Voronoi tessellation.

The realization of Voronoi tessellations for a certain number of expanding nuclei, which is known as the Voronoi foam, can be found in \cite{Icke1987, Weygaert1989}. In the case of two-dimensional realization, the construction of a Voronoi cell consists of the search for all the Delaunay triangles having three nuclei (the center of the circumscribing circle is a vertex of the Voronoi foam). The program proposed by the authors \cite{Icke1987}, allows one to find all the Delaunay triangles having $N_{1}$ as a corner and construct the Voronoi cell belonging to $N_{1}$ by joining the circumcentres of the Delaunay triangles. Having applied this procedure to all nuclei, we obtain the Voronoi tessellation. 

The process of forming the Voronoi tessellation is shown in Fig.~\ref{voronoi-pattern} (left). The points $N_{0}$, $N_{1}$, $N_{2}$ form a Delaunay triangle obtained in a previous search; corresponding Voronoi vertex \textit{V} is shown within the (dashed) circumcircle of $N_{0}$, $N_{1}$, $N_{2}$ as well as stubs of the Voronoi cell walls. On the left hand side of the diagram, the \textit{T} are a sequence of trial points, the third of which produces a circle that encompasses two nuclei, $P_{1}$ and $P_{2}$. The radius of the circumcircle of ($N_{1}$, $N_{2}$, $P_{1}$) being smaller than that of  ($N_{1}$, $N_{2}$, $P_{2}$), the point $P_{2}$ is $N_{3}$, i.e. the third corner of the Delaunay triangle. Thus, the circumcenter of  ($N_{1}$, $N_{2}$, $P_{1}$) is the next Voronoi vertex which, if connected with \textit{V}, produces a complete Voronoi cell wall (\cite{Icke1987}). 

The obtained results could explain the heuristic models that supposing Voronoi tessellations as 3D templates for the galaxy distribution as well as could reproduce a variety of galaxy clustering properties. In an ideal scenario, the LSS is organized by equal spherical voids expanding at the same rate. The walls and filaments would be found precisely between expanding voids, and the resulting LSS web skeleton would the Voronoi tessellation.

The Voronoi tessellation method was picked up and also thrived in our research on a spatial galaxy distribution since 1990-is \cite{Vavilova1995} that allowed us to obtain several priority results. Namely, we elaborated three main approaches in Voronoi tessellation application: (1) to describe a cosmic web skeleton in matter distribution as a Voronoi tessellation with nuclei at low-density regions; (2) to use Voronoi tessellation as a tool for direct measurement of galaxy local concentration and environmental description of low-populated galaxy systems such as triplets, pairs, and isolated galaxies; 3) to apply Voronoi diagrams altogether with machine learning methods for 3D mapping of the Zone of Avoidance of our Galaxy \cite{Vavilova2018, Vavilova2020b}, where Generative Adversarial Network (GAN) algorithms are very useful \cite{Ambrogioni2019, Elyiv2020}. In particular, Coutinho et al. \cite{Coutinho2016} performed verification of various algorithms that can reproduce the cellular structure of the Universe. By comparing the simulated distributions with real observational data, these authors showed that the best algorithm uses the nearest neighbour parameter between galaxies, and that network algorithms can be improved to reproduce the large-scale structure of the Universe.

We give examples in Chapter 2, how manner our developed approach is working. We briefly overview in Chapter 3 various astronomical research with the Voronoi diagrams, accentuating the papers related to the large-scale structure of the Universe, as well as we highlight in Chapter 4 several works and software, where the Voronoi tessellation and machine learning get along well with each other.

\section{Voronoi tessellation of the first, second and third orders: identification of the low-populated galaxy groups, environment effect, and dark matter content}
\label{sec:2}

Because of Voronoi tessellation is a geometrical method based only on galaxy positions, it allows detaching overdensity regions of galaxies in comparison with the background \cite{Vavilova2005b}. We tested it with various samples of galaxies. First of all, we used the Local Supercluster of galaxies, which is well studied among other galaxy superclusters, for identifying galaxy groups of various populations. It was revealed that Voronoi’s tessellation method depends weakly on the richness-parameter of groups, and the number of galaxies in the rich structures is growing rather than in the weak structures with an increase of this parameter \cite{Melnyk2006b}. 

In the first-order Voronoi tessellation, the critical parameter is the volume of the galaxy’s Voronoi cell \textit{V}. This parameter characterizes an environmental galaxy density. The condition of cluster/group membership of a particular galaxy is the relatively small V. This condition is actual when close neighbouring galaxies surround the galaxy. That is why the first order Voronoi tessellation is not corrected for the identification of small isolated galaxy systems \cite{Melnyk2006b}.

\begin{figure}[h]
    \centering
    \includegraphics[width=1.0\textwidth]{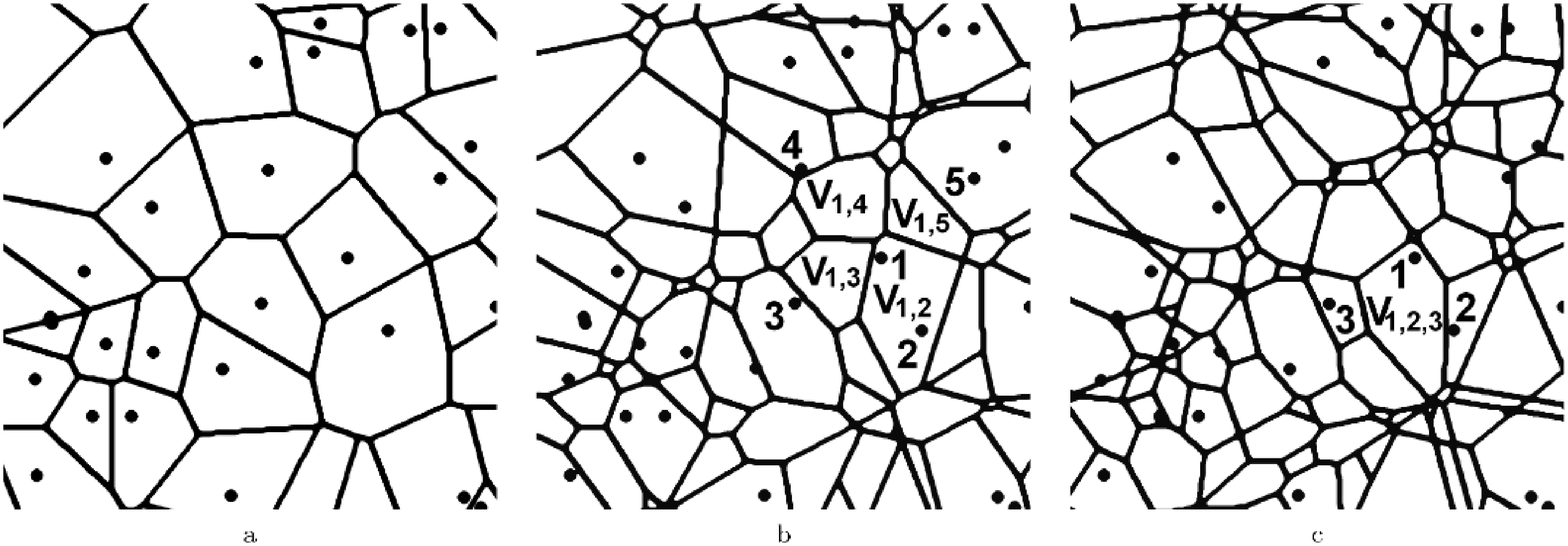}
    \caption{2D Voronoi tessellation of the first- a), second- b) 
    and third- c) order for the same distribution of the random nuclei (\cite{Elyiv2009}, open astronomy).}
    \label{fig:V1}
\end{figure}

We used the second-order Voronoi tessellation for the identification of pairs and single galaxies. Each galaxy $i$ from set $S$ forms the common cells with a certain number of neighbouring galaxies (Fig.~\ref{fig:V1}). So, under neighbouring galaxies of galaxy $i$, we understand only galaxies that create common cells with this galaxy. For example, galaxy 1 creates only 4 common cells ($V_{1,2}$ , $V_{1,3}$ , $V_{1,4}$ , $V_{1,5}$ ) with neighbouring galaxies 2, 3, 4, and 5, respectively. Each pair of galaxies $i$, $j$ is characterized by the dimensionless parameters $p_{i,j}$:

\begin{equation}
  p_{i,j} = \frac{\sqrt[D]{V_{i,j}}}{m_{i,j}}, 
   \label{eq:pij}
\end{equation}

where $D$ -- space dimension, $V_{i,j}$ -- the area (for 2D) or volume (for 3D) of cell, $m_{i,j}$ -- distance between galaxies $i$ and $j$. So, contrary to the first-order tessellation, the second-order
tessellation for set \textit{S} distribution of nuclei is the partition of the space which associates a region $V_{1,2}$ with each pair of nuclei 1 and 2 from $S$ in such a way that all points in $V_{1,2}$ are closer to 1 and 2 than other nuclei from S. Region $V_{1,2}$ is a common cell for nuclei 1 and 2. However, these nuclei do not need to lie in the common cell. For example, nuclei 1 and 5 create the common cell $V_{1,5}$, and they do not lie in this cell. In such a way, the second-order Voronoi tessellation is available for the identification of single galaxies and pairs (Fig.~\ref{fig:V1}b).

\begin{figure}[h]
    \centering
    \includegraphics[width=1.0\textwidth]{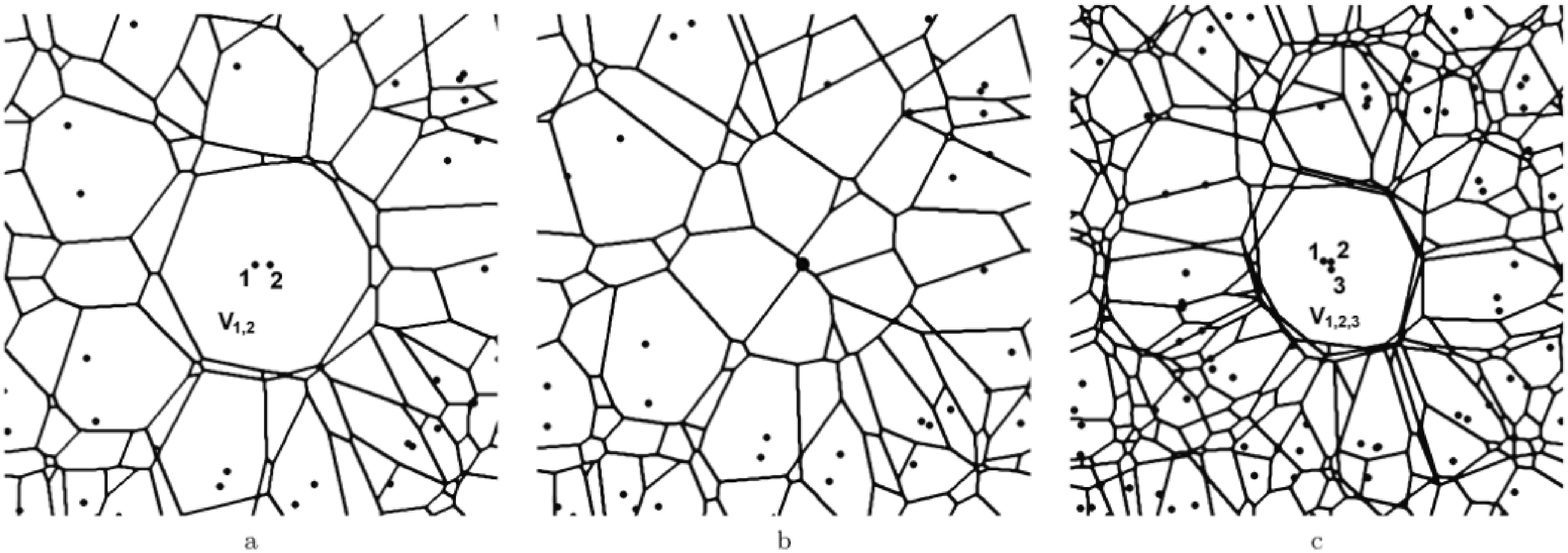}
    \caption{Different configurations of the galaxies: isolated close pair a) and isolated single galaxy b) in the second-order tessellation; isolated close triplet in the third-order tessellation c) (Elyiv,2009, open astronomy).}
    \label{fig:V2}
\end{figure}

Let us introduce the parameter $p_{e}$, which describe only pair environment and does not depend on the distance between pair members directly. We define it as the mean value of $p_{j}(1)$ and $p_l$(2) parameters of the first and second galaxy, excepting $p$ from both sets:

\begin{equation}
   p_{e} =  \frac {\sum_{j = 2}^k p_j (1) + \sum_{l = 2}^n p_l (2)} {k+n-2}, 
   \label{eq:pe}
\end{equation}

where $k$ and $n$ -- number of neighbouring galaxies for 1 and 2 galaxies of geometric pair, respectively. We started sums from $j$ = 2 and $l$ = 2 for excepting $2 · p$, because the first galaxy is neighbour for the second galaxy and vice versa. Therefore $k + n-2$ is sum of neighbouring galaxies of pair members excepting of pair galaxies as neighbouring for each other. Parameter $p_{e}$ depends on the distribution of neighbouring galaxies. A small value of $p_{e}$ points out that such a pair is located in a loose environment. In such case the average volume of common cells of pair components with neighbouring
galaxies is relatively small, and distance between them is significant, see formula ~\eqref{eq:pij} and Fig.~\ref{fig:V2}a.

A single galaxy is a galaxy, which is not a member of any geometric pair. The single galaxies are field galaxies in the environment of geometric pairs. Every single galaxy has the own neighbours; single galaxies and geometric pair members can be among them. According to the second-order Voronoi tessellation, the larger is the degree of galaxy isolation, the larger is the number of neighbours (see Fig.~\ref{fig:V1}b in comparison with Fig.~\ref{fig:V2}b), but these neighbours locate farther. The best parameter that describes the isolation degree of the single galaxy, $s$, is the mean value of all parameters $p_{j}$ of this galaxy:

\begin{equation}
   s =  \frac {\sum_{j = 1}^k p_j} {k} 
   \label{eq:s}
\end{equation}

The third-order Voronoi tessellation is appropriate for the identification of galaxy triplets. It is the partition of the space which associates a region $V_{1,2,3}$ with each triplet of nuclei 1, 2, 3 in such a way that all points in $V_{1,2,3}$ are closer to nuclei 1, 2, 3 than other nuclei from $S$ \cite{Lindenbergh2002}. All points of the common triplet’s cell are closer to galaxies of this triplet than to other galaxies. Similarly to the parameter $p_{i,j}$ for pairs, we can set up the parameter $t_{i,j,u}$ for triplets:

\begin{equation}
  t_{i,j,u} = \frac{\sqrt[D]{V_{i,j,u}}}{max(m_{i,j}, m_{i,u}, m_{j,u}),} 
   \label{eq:tiju}
\end{equation}

where {D} is the space dimension, $V_{i,j,u}$ is the area (for 2D) or volume (for 3D) of the cell, and $m_{i,j}$, $m_{i,u}$, $m_{j,u}$ are the distances between galaxies in the triplet. A geometric triplet in the third-order Voronoi tessellation contains three galaxies that have a common cell and the same maximal
parameters $t_{max}(1) = t_{max}(2) = t_{max}(3) = t$. The parameter $t$ characterizes a degree of geometric triplet isolation. We can define the parameter of triplet environment $t_{e}$ as the mean value of parameters $t_{i}(1)$, $t_{j}(2)$, and $t_{u}(3)$, except $t$ from three sets:

\begin{equation}
   t_{e} =  \frac {\sum_{i = 2}^k t_i (1) + \sum_{j = 2}^n t_j (2) + \sum_{u = 2}^q t_u (3)} {k+n+q-3} 
   \label{eq:te}
\end{equation}

here in the case of the third-order Voronoi tessellation, $k$, $n$, and $q$ denote the number of neighbouring triplets which contain galaxies 1, 2, and 3, respectively. Therefore, ($k$ + $n$ + $q$ - 3) is the number of neighbouring triplets for a certain triplet that contain at least one galaxy from this triplet (see, Fig.~\ref{fig:V2}).

Parameters $p$, $s$, and $t$ are the basic ones and define the isolation degree of a galaxy pair, single galaxy, or triplet compared to the background, respectively. Parameters $p_e$ and $t_e$ are additional ones and contain information about the distribution of the neighbouring galaxies (environment).
Similar to the second- and third-order Voronoi tessellations, it is possible to apply more high-order Voronoi tessellations to identify galaxy quartets and quintets, etc.

So, one can use galaxies as the nuclei of the Voronoi tessellation taking into account equatorial
coordinates $\alpha$, $\delta$ and radial velocities of galaxies $V_h$ only. For the construction of the 3D Voronoi tessellations, it is necessary to determine the distances in 3D space. The spatial distance between two galaxies can be decomposed into projected (tangential) distance r and radial component v (difference of the radial velocities). We can determine the projected distance with a relatively high accuracy. Simultaneously, the radial component has errors due to the inaccuracy of radial velocity measurement of each galaxy and existing strong peculiar velocities (due to virial motions of galaxies in groups and clusters).
As a result, the galaxy distribution in the radial velocities space is extended along the radial component, the so-called fingers-of-God effect. This is attributed to the random velocity dispersion in a
galaxy volume-limited sample that cause a galaxy’s velocity to deviate from pure Hubble flow, stretching out a group of galaxies in redshift space (\cite{Jackson1972, Melnyk2009}). Various authors take into account this effect in their way, depending on the specifics of their problem. For example, Marinoni et al. (\cite{Marinoni2002}) chose some cylindrical window of clustering, which is extended along the radial component. We introduced the weight for a radial component (\cite{Elyiv2009}, avoiding the problem of tangential and radial distance in equivalence to apply the high-order 3D Voronoi tessellation method.

An efficient way to show Voronoi tessellation advantages was to apply it to the galaxy samples from the Local Supercluster \cite{Vavilova2005c, Melnyk2006b, Vavilova2015} and the Sloan Digital Sky Survey (SDSS), where at the first time we examined it for spectroscopic aims \cite{Melnyk2006c, Elyiv2009, Vavilova2009, Melnyk2012, OMill2012, Pulatova2015}. We did not consider galaxies that located within $2^{o}$ near borders, because the correct estimation of Voronoi cell volume is not possible in this case.
Selecting single galaxies and pairs by the second-order Voronoi tessellation, as well as triplets by the third-order Voronoi tessellation method, we obtained 2196 geometric pairs, 1182 triplets and 2394 single galaxies. We did not make a clear division between physical gravitationally bound systems and
non-physical ones, following the supposition that the more isolated a system is, the higher probability that it is physical (compact pairs are with $R_h$ < 150 kpc and triplets are with $R_h$ < 200 kpc). 

\begin{figure}[h]
    \centering
    \includegraphics[width=1.0\textwidth]{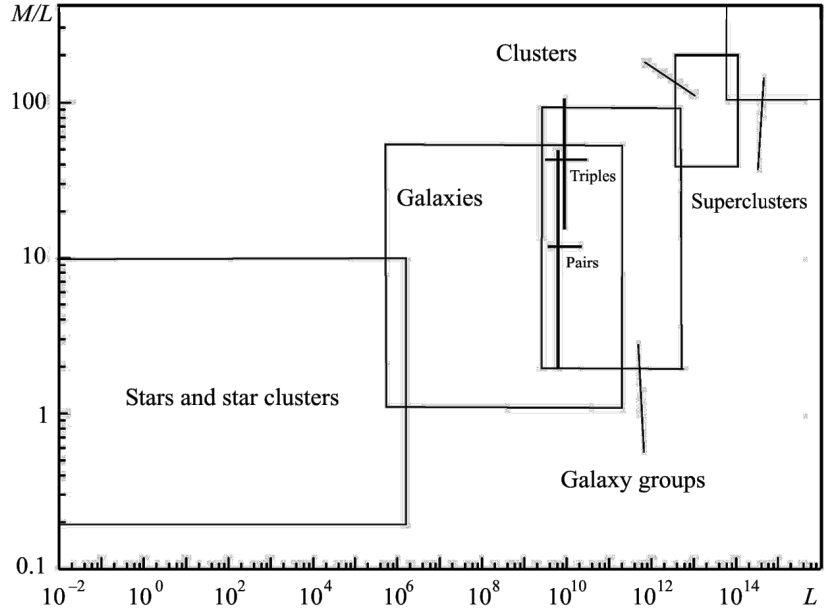}
    \caption{Mass-to-luminosity ratio diagram for galaxy systems of different population (star clusters, galaxies, galaxy groups, clusters and superclusters), where the result for the low-populated groups (Melnyk, 2009) is pointed (Vavilova, 2015, open astronomy).}
    \label{fig:DM}
\end{figure}

\begin{figure}[h]
    \centering
    \includegraphics[width=1.0\textwidth]{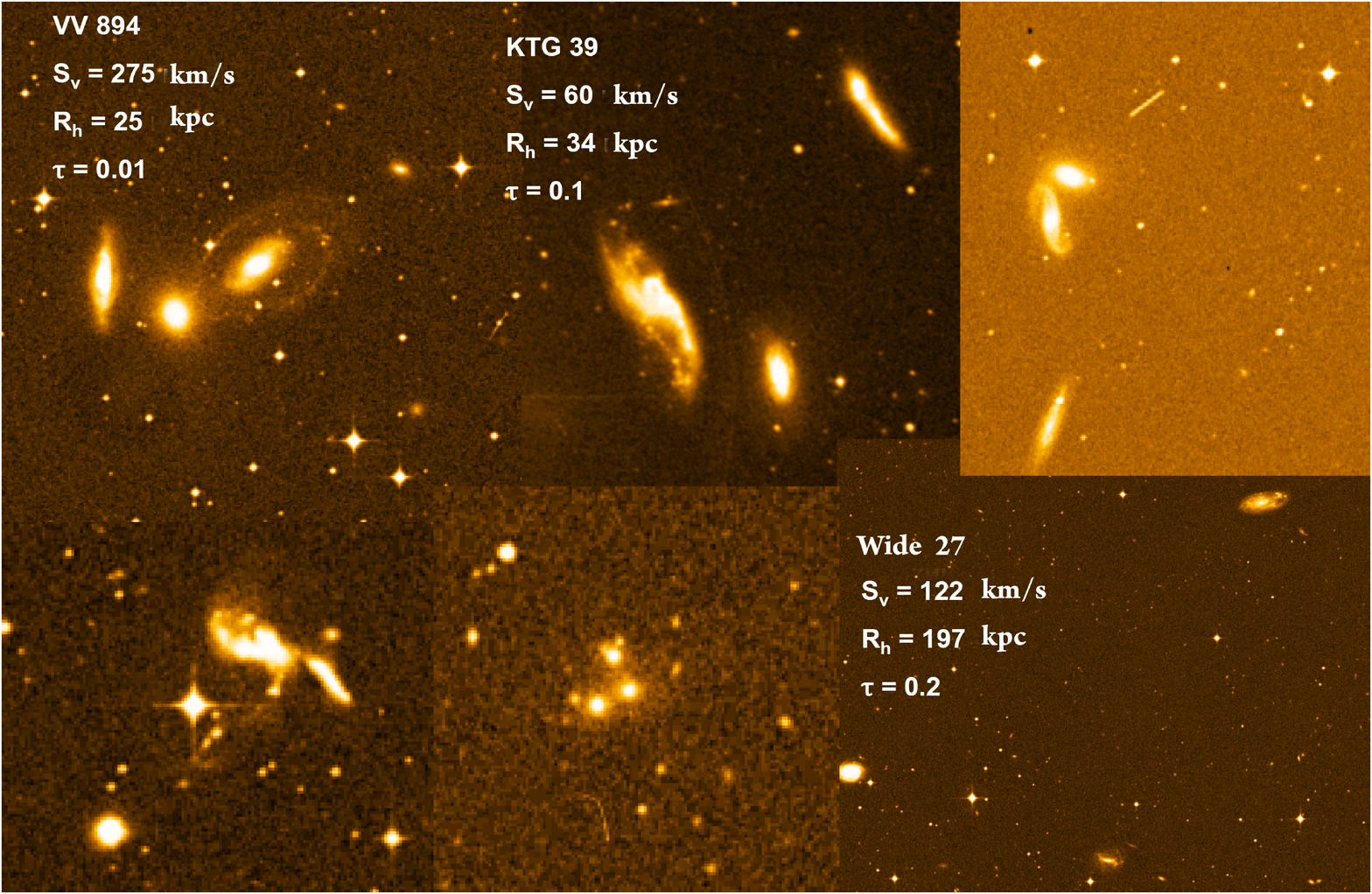}
    \caption{The interacting (VV894), most compact (KTG39), and wide triplets of galaxies, where $S_{v}$ is the rms velocity of galaxies with respect to the triplet centre, $R_{h}$ -- harmonic mean radii of the triplet, $\tau = 2H_{0}R_{h}/S_{v}$ -- its dimensionless crossing time (Vavilova, 2015, open astronomy).}
    \label{fig:trio}
\end{figure}

Estimating the dark matter content in the low-populated groups, we obtained the median values of mass-to-luminosity ratio (\cite{Elyiv2009}):  $12M_{solar}/L_{solar}$ for the isolated pairs and $44M_{solar}/L_{solar}$ for the isolated triplets. Note that for the most compact pairs and triplets (with R < 50 (100) kpc, respectively) there is not a very large difference in dark matter content for pairs and triplets: $7~M_{solar}/L_{solar}$ and $8~M_{solar}/L_{solar}$. The mass-to-luminosity ratio diagram for galaxy systems of different population (star clusters, galaxies, galaxy groups, including the low-populated ones, clusters, and superclusters) is presented in Fig.~\ref{fig:DM}. Several examples of isolated triplets of galaxies are given in Fig.~\ref{fig:trio}. We conclude about the dark matter distribution that for the dynamically younger sparsely groups (triplets), dark matter is more likely associated with the individual galaxy halos, for the interacting and late sparsely groups the dark matter lies in a common halo of galaxy groups.

Using an inverse volume of Voronoi cell (1/$V$) as a parameter describing the local environmental density of a galaxy, we considered the volume-limited SDSS (DR5 and DR9) galaxy samples ($0.02<z<0.1$, $-24<M_r<-19.4$)  \cite{Melnyk2012, Dobrycheva2013, Dobrycheva2015, Dobrycheva2017} and found that
\begin{itemize}
  \item the early type galaxies prefer to reside in the Voronoi cells of smaller volumes (i.e., dense environments) than the late type galaxies, which are located in the larger Voronoi cells (i.e., sparse environments);
  \item the relationships between the morphological types and the $u-r$, $g-i$, and $r-z$ color indices of pairs of galaxies with radial velocities $3000<V<9500$ km/s evident that the Holmberg effect is not revealing, by the other words, it can be considered only in historical aspect \cite{Dobrycheva2016};
  \item properties of such small groups as pairs and triplets, where segregation by luminosity was clearly observed, are fit well to Dressler effect: galaxies in isolated pairs and triplets are on average two times more luminous than isolated galaxies;
  \item the dependence of the color indices and stellar magnitudes is effective for the automated morphological classification of the galaxies ($E$ -- early types, $L$ -- late types).
\end{itemize}

The morphological types of the galaxies were divided into two classes: Early - $E$ (from elliptical and lenticular) and  Late - $L$ (from $Sa$ to $Irr$). The absolute magnitude 

\begin{equation}
   M_{r} =  m_{r} - 5log(V/H_{0})-25-K(z)-ext
   \label{eq:Mr}
\end{equation}

could be corrected for Galactic absorption $ext$ in accordance with ~\cite{Schlegel1998} and $K$ - correction $K(z)$ according to ~\cite{Chilingarian2010}. Here we used the CDM model of the Universe with the WMPS7 cosmological parameters ($\Omega_{M}$ = 0.27, $\Omega_{\Lambda}$ = 0.73, $\Omega_{k}$ = 0, $H_{0}$ = 0.71). In order to apply the Voronoi tessellation method we should done transition from equatorial coordinates and velocities to the comoving $x$, $y$, $z$ coordinates for each central galaxy in the sample ($M_r <$ -20.7). To do this we can transform the redshift $z$ to the corresponding distance $\chi(z)$ for each galaxy by integrating as follows

\begin{equation}
     \chi(z) =  D_{H}\int_{0}^{z}  \frac {dz'}{E(z')} 
   \label{eq:hi(z)}
\end{equation}

where $D_{H}=c/H_{0}$ is the Hubble distance and $E(z')$ is the Hubble parameter, defined as follows

\begin{equation}
     E(z') = \sqrt{\Omega_{M}(1+z)^3 + \Omega_{k}(1+z)^2 + \Omega_{\Lambda}} 
   \label{eq:E(z)}
\end{equation}

The coordinates $x$, $y$, $z$ of the galaxies in the comoving space are determined as follows

\begin{equation}
   x = \chi(z)cos(\theta)cos(\phi)
   \label{eq:x}
\end{equation}

where ($\theta$) is the declination of each galaxy, ($\phi$) is the right ascension, and ($\chi(z)$) is the corresponding distance for redshift $z$. After getting the three-dimensional Cartesian coordinates of the galaxies, we divided the geometrical space occupied by the galaxy sample in mosaic cells (volumes $V$ in the 3D case). Each cell has a galaxy as a nucleus and consists of elementary volumes of space closer to this galaxy than to any other galaxy \cite{Matsuda1984}. The use of the Voronoi tessellation to isolate groups of galaxies in three dimensions has been described in detail by Melnyk et al. \cite{Melnyk2006b}. Fig.~\ref{fig:V1}a shows an example of the Voronoi tessellation in a two dimensional case to make it easier to see. Let us use the value of inverse volume ($1/V$) of the Voronoi cells to describe the density of galaxy environments; when $1/V$ is higher, a galaxy is less isolated.

\begin{figure}[h!]
\centering
    \includegraphics[width=.95\textwidth]{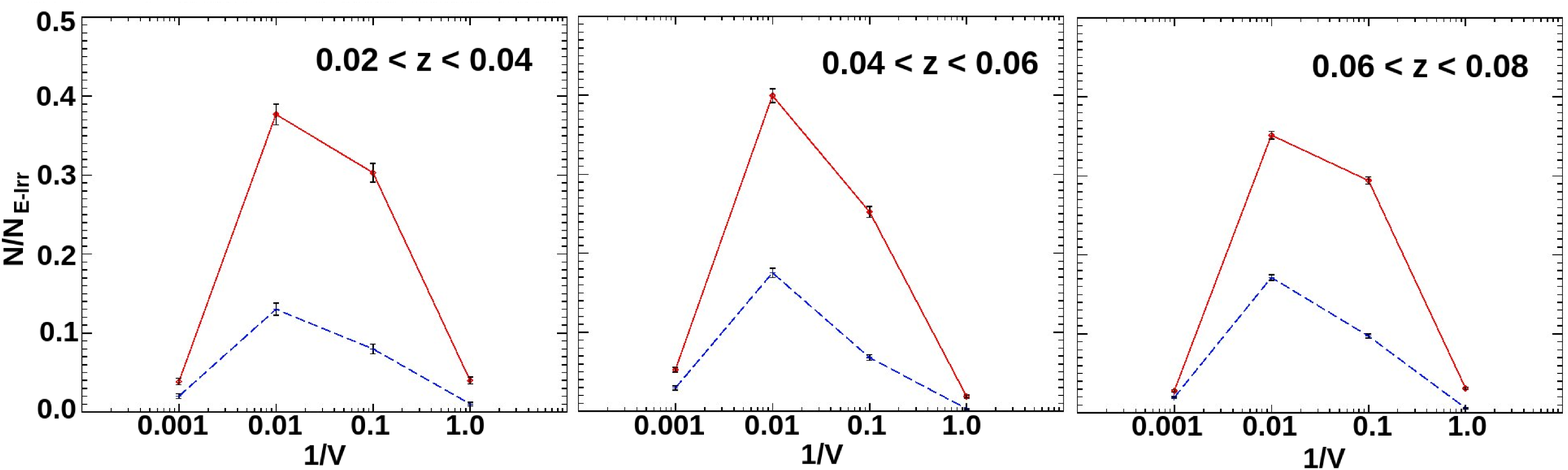}
\caption{The distribution of the number of galaxies vs inverse volume of the Voronoi cell (local density parameter), with early morphological type $E$ indicated by red lines and late type $L$ indicated by blue lines, for different ranges of redshift; absolute stellar magnitude of galaxies selected from the SDSS at z < 0.1 is $-22.5<M_{r}<-21.5$. The number of galaxies in each bin is normalized to the total number of $E \div L$ within the given subsample. The number of central bright $E$ and $L$ galaxies is as follows: $E$ = 1636, $L$ = 459 for $0.02<z<0.04$, $E$ = 3609, $L$ = 1247 for $0.04<z<0.06$, $E$ = 9432, $L$ = 3596 for $0.06<z<0.08$ (Dobrycheva, 2015).}
\label{fig6}
\end{figure}

Examples of the distributions of $E$ and $L$ galaxies vs. inverse volume of the Voronoi cells that contain them are shown in Fig.~\ref{fig6}. In work \cite{Dobrycheva2015} we grouped galaxies from the SDDSS sample at z < 0.1 into 4 logarithmic intervals $1/V<0.001$, $0.001<1/V<0.01$, $0.01<1/V<0.1$, and $1/V>0.1$ for four ranges of the redshift $0.02<z<0.04$, $0.04<z<0.06$, $0.06<z<0.08$, and $0.08<z<0.1$ (in the rows) and for different ranges of absolute stellar magnitude, $-21.5<M_{r}<-20.7$, $-22.5<M_{r}<-21.5$, and $M_{r}<-22.5$. The number of galaxies in each bin for the $E$ and $L$ types is normalized to the total number of $E \div L$ galaxies within the given subsample. Fig.~\ref{fig6} shows that the fraction of galaxies of spiral and late types becomes larger while redshift increasing, while the fraction of early types, on the contrary, is smaller. That follows the well known evolutionary trend of a reduction in the number of galaxies with suppression of star formation for increasing redshift \cite{Cucciati2006, Tal2014}, even at comparatively low redshifts down to $z<0.1$. Also, for the brighter galaxies in the sample, the fraction of galaxies of earlier types is larger since, on the average, earlier types have higher luminosities (the well-known morphological type vs. colour indices/luminosity relation) \cite{Blanton2005, Park2007, Hogg2004}. The brightest galaxies of earlier types with $M_{r}<-22.5$ appear preferentially in denser environments: the peak of the distribution of the inverse volumes of the Voronoi cells for the $E$ types lie within the interval $0.01<1/V<0.1$, while in other intervals of $M_{r}$, for the $L$ types the peak of the distribution always is within $0.001<1/V<0.01$ (the morphology-density relation \cite{Dressler1980, Blanton2005, Hogg2004, Dobrycheva2016}).

We can also determine the density of galaxies in a Voronoi cell, including their faint satellites, i.e., galaxies with $M_{r}r>-20.7$: $(n+1)/V$, where n is the number of faint galaxies in the Voronoi cell, and $V$ is the volume of the Voronoi cell. We also constructed distributions of early $E$ and $L$ types galaxies in dependence on the parameter $(n+1)/V$ in four intervals: $(n+1)/V<0.01$, $0.01<(n+1)/V<0.1$, $0.1<(n+1)/V<1$, and $(n+1)/V>1$. The number of galaxies is normalized to the number of $E \div L$ galaxies within the given range of $(n+1)/V$. We examined the density of galaxies only in the first two redshift intervals, since we cannot evaluate the evolution of their properties at a higher $z$ because there are not enough faint galaxies. However, we can compare the galaxies' environmental density as a function of the absolute magnitude and morphological type of the bright central galaxy. Thus, the fraction of early types of central galaxies increases with increasing environmental density, while, on the other hand, the fraction of late types decreases; that is, the earlier types are in a denser environment than the late types. When the central galaxy is brighter, the fraction of early types in a subsample will be larger \cite{Dobrycheva2015, Vavilova2020c}.

\section{The Voronoi tessellation in astrophysical research}
\label{sec:3}

Ebeling and Wiedenmann \cite{Ebeling1993} were the first to apply the Voronoi tessellation for finding galaxy groups and clusters. Later such an approach was used by Ramella et al. \cite{Ramella2001}, Kim et al. \cite{Kim2002}, Lopes et al. \cite{Lopes2004}, Barrena et al. \cite{Barrena2005}, Melnyk et al. \cite{Melnyk2006c}, Panko and Flin \cite{Panko2006}. Doroshkevich \cite{Doroshkevich1997} introduced its for filaments and walls (1D and 2D LSS structures) as well as Neyrinck \cite{Neyrinck2008} for the search of voids in a spatial galaxy distribution. 

We note some important earlier works as concerns with other applications of Voronoi diagrams to the large-scale galaxy distribution: for revealing the quasi-periodicities in a pencil-beam survey ~\cite{Subba1992, Ickeuchi1991}, for a description of constraints on the Voronoi model when applied to the isotropic cosmic microwave background ~\cite{Coles1990}. A significant contribution for Voronoi tessellation application to various astronomical tasks was made by Zanninetti, who considered two- and three-dimensional cases of the explosion scenario likely supernova events and developed a dynamical method allowing to describe the explosion phases ~\cite{Zaninetti1989, Zaninetti1990}.

Ramella et al. ~\cite{Ramella2001} created a Voronoi Galaxy Cluster Finder, which uses positions and magnitudes of galaxies to define galaxy clusters and extract its parameters: size, richness, central density, etc. The 3D Voronoi tessellation for galaxy group identification was realized by Marinoni et al. \cite{Marinoni2002} and Cooper et al. \cite{Cooper2005}. Weygaert et al. prepared a useful review of the spatial galaxy distribution and Delaunay and Voronoi tessellations \cite{Weygaert2009, Hidding2015}. They discussed the Delaunay Tessellation Field Estimator (DTFE) and the concept of Alphashapes for matter distribution; the Multiscale Morphology Filter (MMF), which uses the DTFE for detachment of filaments, sheets, and clusters; the Watershed Voidfinder (WVF) to identify voids.

The era of big data surveys (see, for example, review in work by Vavilova et al. \cite{Vavilova2020a} accelerated the Voronoi diagrams application on a spatial galaxy distribution properties and environment influence: $z = 0.1 - 3.0$, COSMOS survey \cite{Scoville2013}; z {\ensuremath{\leq}} 0.5, Herschel-ATLAS/GAMA \cite{Burton2013}; z < 0.1, Coma Supercluster \cite{Cubul2014}; z < 0.3, ALHAMBRA survey \cite{SanRoman2018}. S{\"o}chting et al. used Voronoi tessellation within overlapping slices in the photometric redshift space (0.2<z<3.0). It allowed them to detach region $z\sim0.4$ with a slow emergence of virialized clusters accordingly to the hierarchical scenario and to detect new superclusters as the peaks of a matter distribution up to z = 2.9 \cite{Sochting2012}. As for the Voronoi tessellation cluster finder algorithms, we note the work by Soares et al., who developed it to produce reliable cluster catalogs up to $z=1$ or beyond and down to $10^{13.5}$ solar masses. They built the Voronoi tessellation cluster finder in photometric redshift shells and used the two-point correlation function of the galaxies in the field to determine the density threshold for the detection of cluster candidates and to establish their significance \cite{Soares2011}. 

A principal new galaxy cluster finder based on a 3D Voronoi Tessellation plus a maximum likelihood estimator, followed by gapping-filtering in radial velocity ($VoML+G$), was developed by Pereira et al. \cite{Pereira2017a, Pereira2017b}. They applied it successfully to find optical clusters ($R_{200}$) in the Local Universe as well as Santiago-Batista et al. for the identification of continuous filaments in the environment of superclusters \cite{SantiagoBautista2020}. Grokhovskaya et al. developed filtering algorithms of multiparameter analysis of the large scale distribution of galaxies (identification of galaxy systems and voids) in narrow slices in the entire range of redshifts of HS 47.5-22 constructing density contrast maps, namely with adaptive kernel and Voronoi tessellation \cite{Grokhovskaya2019}. The 3D Voronoi tessellation application to the DEEP2 survey was first introduced by Gerke et al. \cite{Gerke2005}. Meanwhile, Shen Ying et al. \cite{Ying2015} proposed an algorithm which computes the cluster of 3D points by applying a set of 3D Voronoi cells and allows a 3D point cluster pattern can be highlighted and easily recognized.

Hung et al. have demonstrated that Voronoi tessellation Monte-Carlo mapping is beneficial for studying the environment effect on galaxy evolution in high-redshift large-scale structures (z$\sim$1) in the ORELSE survey (Observations of Redshift Evolution in Large Scale Environments) \cite{Hung2019}. An exciting application of Voronoi tessellation was proposed by Lam et al. \cite{Lam2019}: for constructing the white dwarfs luminosity functions they used parameters of proper motion and colours from the Pan-STARRS\,1\,3$\pi$ Steradian Survey Processing Version 2; for improving the accuracy of the maximum volume method they used Voronoi tessellation space binning to recalculate photometric/astrometric uncertainties. It helped to estimate disk-to=halo dark matter ratio as 100. Another a non-parametric method for estimating
halo concentration using Voronoi tessellation, TesseRACT, was proposed by Lang et al. \cite{Lang2015}, who showed that it fit well with non-spherical halos and more accurate at recovering intermediate concentrations for N-body halos than techniques that assume spherical symmetry.

\begin{figure}[h!]
\centering
    \includegraphics[width=.95\textwidth]{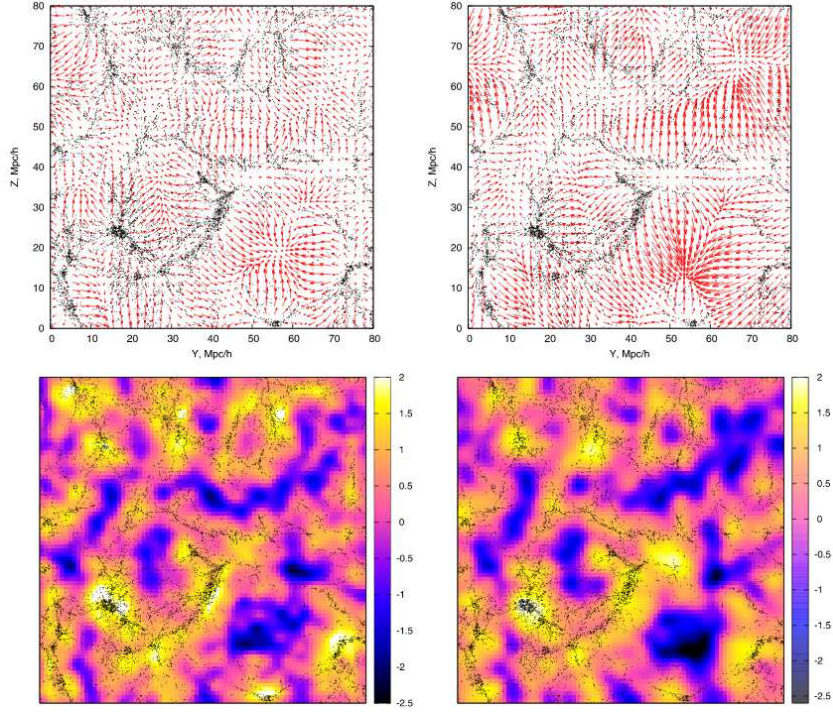}
\caption{The reconstructed displacement field  (top panels) and its divergence (bottom panels) obtained with the two void finders, the Uncorrelating Void Finder (left-hand panels) and the Lagrangian Zel’dovich Void Finder (right-hand panels). }
\label{fig7}
\end{figure}

The very interesting algorithm, Void Finder ZOBOV (ZOnes Bordering On Voidness), based on Voronoi tessellation, was proposed by Neyrinck et al. \cite{Neyrinck2008}. This algorithm finds density depressions galaxy distribution without free parameters. To estimate local density, it uses the Voronoi tessellation. One of the output of this algorithm is the probability that each void arises from Poisson fluctuations. However, Elyiv et al. \cite{Elyiv2015} have demonstrated a weak spot for ZOBOV void finder. Voids are the lowest density regions, so any method that uses the positions of galaxies directly to measure density for identifying the voids is then prone to shot noise error since voids are the regions with a very low concentration of galaxies by definition (Fig. \ref{fig7}).
The Void IDentification and Examination toolkit (VIDE) developed by Sutter et al. \cite{Sutter2015} includes the parameter-free void finder ZOBOV, where ``Voronoi tessellation of the tracer particles is used to estimate the density field followed by a watershed algorithm to group Voronoi cells into zones and subsequently voids''. 

Zaninetti in series of works \cite{Zaninetti2010a, Zaninetti2012} developed a practical statistics for the voids between galaxies with two new survival functions and considered the 3D distribution of the volumes of Poissonian Voronoi Diagrams to their 2D cross-sections in the assumption of gamma-function for the 3D statistics of the volumes of the voids in the Local Universe. He also conducted simulations \cite{Zaninetti2010b} of a spatial galaxy distribution using the Poissonian Voronoi polyhedra and the 2dF Galaxy Redshift Survey and the Third Reference Catalog of Bright Galaxies; Zaninetti gives a brief overview of a current status of the research on the statistics of the Voronoi Diagrams in \cite{Zaninetti2013}.

Among other astronomical tasks, the Voronoi diagrams have been used for image processing, adaptive smoothing, segmentation, for signal-to-noise ratio balancing ~\cite{Chadha2016}, for spatial structure of the solar wind and solar-terrestrial connections \cite{Borovsky2018}, for spectrography data analysis in different electromagnetic regions \cite{Cappellari2002, Cappellari2003, Diehl2006}, in the moving-mesh cosmology simulation ~\cite{Springel2010} and \cite{Weinberger2020} (AREPO Public Code), chemical evolution in the early universe \cite{Chiaki2016}, star formation simulation \cite{Hubber2016}, spatial distribution of lunar craters \cite{Honda2019}. For example, Cabrera et al. ~\cite{Cabrera2008} applied the Voronoi diagram for image reconstruction technique in the interferometric data based on the Bayesian approach. Cadha et al. proposed Voronoi compact image descriptors and showed that Voronoi partitioning improves the geometric invariance and performance of image retrieval ~\cite{Chadha2016} as well as they developed a Voronoi-based machine learning method (deep convolution neural network). As for the cosmological simulation, Busch and White \cite{Busch2020} used Voronoi tessellation for a hierarchical tree structure that allowed them to associate local density peaks with disjoint subsets of particles and to analyze mass distribution at different levels of threshold. Similar to our work \cite{Dobrycheva2015}, when we introduced parameter of the volume of Voronoi cell to study environment influence on galaxies from the SDSS, Paranjape \& Alam \cite{Paranjape2020} applied inverse local number density parameter to study 
physical effects for such properties as halo (subhalo) mass, large-scale environment, etc. in various cosmological dark matter models and concluded that the Voronoi volume function gives a new mathematical instrument for galaxy evolution physics and dark sector study.

Neyrinck developed the sectional-Voronoi algorithms in Python for cosmic-web research, because the Voronoi/Delaunay duals and origami tessellation give a wide class of spiderwebs. ``Voronoi edges are perpendicular bisectors of their corresponding Delaunay edges; the `bisector' part can be relaxed. Each Voronoi edge may be slid along its Delaunay edge, closer to one of the generators. They may not be slid entirely independently, though, since the Voronoi edges must still join vertices. There turns out to be one extra degree of freedom per generator, causing its cell to expand or contract. The result is a sectional-Voronoi diagram, a section through a higher-dimensional Voronoi tessellation. A generator’s extra degree of freedom in a sectional-Voronoi diagram can be thought of as its distance from the space being tessellated. A sectional-Voronoi diagram can also be thought of as a Voronoi tessellation in which each generator may have a different additive `power' in the distance function used to determine which points are closest to the generator (thus an alternative term, power diagram). Ash and Bolker \cite{Ash1986} showed that 2D spiderwebs and sectional Voronoi tessellations are equivalent'' (cited by \cite{Neyrinck2008b}. The package is available at \url{https://github.com/neyrinck/sectional-tess}, \url{https://mybinder.org/v2/gh/neyrinck/sectional-tess/master}.

In the present day, the Voronoi diagrams methods have many applications in various fields of science and technology, as well as in social sciences and visual art ~\cite{Aurenhammer1991, Aurenhammer2000}. They are commonly used in computational fluid dynamics, computational geometry, geolocation and logistics, game dev programming, cartography, engineering, liquid crystal electronic technology, etc. For the first time, the Voronoi tessellation was utilized by Debnath et al. \cite{Debnath2015} for the discoveries in the particle physics beyond the Standard Model at the Large Hadron Collider at CERN. ``Since such tessellations capture the underlying probability distributions of the phase space, interesting features in the data can be detected by studying the geometrical aspects of the ensemble of Voronoi cells (cited by \cite{Debnath2018}). These methods allow identifying kinematic edges in two dimensions and generalize the technique for robust detection of phase space boundaries, which could be applied to discover new physics.
An interesting library of ''Voronoi Diagrams: Applications from Archaeology to Zoology" is collected by Scot Drysdale on the website \url{https://www.ics.uci.edu/~eppstein/gina/scot.drysdale.html}.

\section{The Voronoi tessellation and Machine learning}
\label{sec:4}

Straight application of classical Voronoi diagram in Machine Learning is the k-nearest neighbors (k-NN) algorithm at the number of neighbors $k=1$. In the case of the classifier, the output class is choosing among its \textit{k} the closest neighbors. Each of them gives a contribution to the class with some weight. Normally weight is inverse to the distance between target object and neighbor (closer neighbors will have a stronger influence than further neighbors) or uniform (all points in neighborhood are weighted equally). If $k=1$, then the object is just linked to the class of the nearest neighbor. From the other side, it could be interpreted as the building of the Voronoi diagram by training objects as nuclei of the diagram. The target object will have a class depending on which Voronoi cell it resides. Bring your data to life.

A set of programming codes for 1-NN visualization ($k=1$) with examples (Hover Voronoi, a demonstration of d3-Delaunay, Voronoi Labels, Voronoi neighbors, Voronoi update) are available on the website \url{https://observablehq.com/@d3/} by Mike Bostock (2018). For the color image segmentation problem in computer vision, an adaptive and unsupervised clustering approach with Voronoi diagrams was introduced, which outperforms the existing algorithms \cite{Hettiarachchi2016}. A Python library ''Pycobra" contains several ensemble machine learning algorithms and visualization tools based on the Voronoi tessellations \cite{Guedj2017}. It can be downloaded from the Python Package Index (PyPi) and Machine Learning Open Source Software (MLOSS) at \url{https://github.com/bhargavvader/pycobra}.

In the case of $k>1$, we should use the concept of high order Voronoi diagram, where a cell represents the set of points in space closer to a given $k$ nuclei that to all others (see, Chapter 2 and works by Elyiv et al. \cite{Elyiv2009}, \cite{Vavilova2020b}). In this case, k--order Voronoi space dividing can help us to find k-near neighbors directly. The crossing of high-order Voronoi diagram borders represent changing the set of k near neighbors. In k--NN regression, the output value for the target object is the average of the values of k nearest neighbors. If each neighbor has equal weight, it means that for each cell could be assigned pre-calculated averaged value. Next, if the target object resides in this cell, automatic assigned could be done. In all these cases, creating a Voronoi diagram on the training sample could make a faster k--NN algorithm application. 

For example, Inkulu and Kapoor \cite{Inkulu2011} presented an algorithm covering the Voronoi diagram with hyperboxes, which provides ANN queries. Another parallel spatial range query algorithm based on Voronoi diagrams and MR-tree, which is benefiting from the k-NN, is developed by Fu and Liu \cite{Fu2012}.

Voronoi diagram also has a wide application in deep learning. In work \cite{Balestriero2019}, the authors studied the geometry of Deep Artificial Neural Networks with piecewise affine and convex nonlinearities. The authors demonstrated that each layer's input space partition corresponds to the Voronoi diagram with several regions that grow exponentially with increasing neurons. Numerical experiments for classification problems support their main theoretical results are expressed by the Deep ANN decision boundary in the input space, a measure of its curvature that depends on the network architecture, activation functions, and weights. In work \cite{Igashov2020} the authors presented
a Deep Convolution Neural Network (CNN) constructed on a Voronoi tessellation of 3D molecular structures of proteins (VoroCNN model). Both convolution and pooling operations were used as a part of network architecture to predict local qualities of 3D protein folds.
They computed Voronoi tessellation of molecular 3D structures and converted them into a protein interaction graph. The graph's critical property is that it implicitly keeps the information about the spatial relationship between the atoms of the protein model. The authors claim that for presently available amounts of data and computational resources, Voronoi tessellation is the best representation of 3D protein structure than raw volumetric data.

\section{Instead of Conclusion}
\label{sec:5}
Today, hierarchical clustering is a common scenario for the evolution of galaxies. The fact that galaxies are observed mostly at redshifts to $z \sim 5$, while the most distant observed galaxy clusters are at $z \sim 2$, suggests that galaxies and sparsely populated groups were formed first, and galaxy clusters later by subcluster merging and/or
via capturing galaxies and galaxy groups. The hierarchical clustering scenario is in good agreement with the cosmological $\Lambda$CDM model. Having great success in explaining the formation of the large-scale structure of the Universe as a whole, this model faces potentially severe problems on the scales of individual dark halo of galaxies and galaxy clusters, with statistics of the distribution of galaxies with different morphological types in a wide range of redshifts, with evolutionary properties of sparsely populated groups and galaxy clusters, with the lack of data on the large-scale structure of the Universe behind the Zone of Avoidance of the Galaxy.

In this context, we have demonstrated the perfection and elegance of the Voronoi tessellation in solving many astronomical problems, focusing on its effectiveness for describing the web of large-scale structures of the Universe and data mining of its properties at various redshifts from early epochs to the scales of the Local Universe.

\begin{figure}
    \centering
    \includegraphics[width=0.8\textwidth]{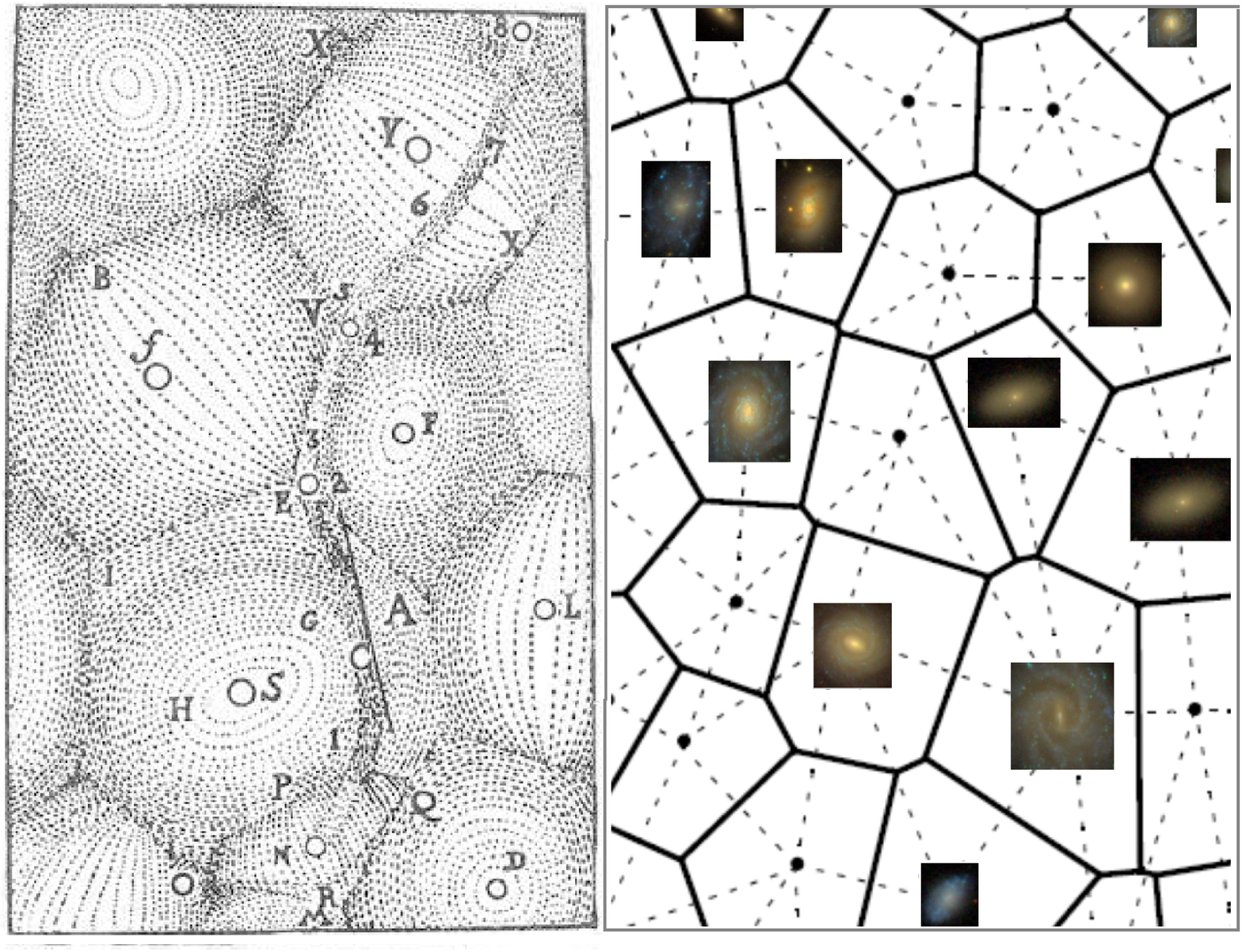}
        \caption{(Left) Vortex theory applied to the Solar system (R. Descartes, 1644) (Aurenhammer, 2000, open access). (Right) Illustration of the Voronoi tessellation for galaxy web distribution.}
    \label{fig:Dec1644}
\end{figure}

``God first partitioned the plenum into equal-sized portions, and then placed these bodies into various circular motions that, ultimately, formed the three elements of matter and the vortex systems'' (cited by R. Descartes, 1644 year \cite{Descartes1644}, vol.III, article 46, in \cite{Aurenhammer2000}). ''The modern view shoves baryogenesis, leptogenesis, WIMP-- genesis, and all very far back in time, but builds up structure continuously, using not-very-special initial conditions and gravity (plus perhaps other forces) to develop what we see today. In between come some remarkable constructs, including Thomas Wright’s hierarchy, Descartes’s Voronoi tessellation of whirlpools in the ether, Alfred Russel Wallace’s (yes, the evolution guy) ''Goldilocks” location for the Solar system, Cornelis Easton’s off-center spiral arms, and the Kapteyn Universe" (cited by V.\,Trimble, 2014 year \cite{Trimble2014}). We have combined this representation, which is consonant with ours, in Fig. \ref{fig:Dec1644} as an illustration of partitioning the space into cells for the subsequent extraction of the physical essence of the phenomena: one of them displays classical physics, Vortex theory applied to the Solar system (Descartes, 1644), the other gives a visualization of galaxy distribution through the 2D- Voronoi tessellation. 

\begin{acknowledgement}
This work was partially conducted in the frame of the budgetary program of the NAS of Ukraine ``Support for the development of priority fields of scientific research'' (CPCEL 6541230).
\end{acknowledgement}



\begin{thebibliography}{99.}

%
%
%
%
%
%
%
%
%
%

\bibitem{Ambrogioni2019} Ambrogioni, L.,G{\"u}{\c{c}}l{\"u}, U., van Gerven, M.: k-GANs: Ensemble of Generative Models with Semi-Discrete Optimal Transport. arXiv:1907.04050 (2019).

\bibitem{Ash1986} Ash, P.F., Bolker, E.D. Generalized Dirichlet tessellations. Geometriae Dedicata. \textbf{20(2)}, 209–243 (1986).

\bibitem{Aurenhammer1991}  Aurenhammer, F.: Voronoi Diagrams. A Survey of a Fundamental Geometric Data Structure. ACM Computing Surveys. \textbf{23(3)}, 345-405 (1991).

\bibitem{Aurenhammer2000}  Aurenhammer F., Klein R.: In Voronoi diagrams. Ed. Sack J-R. Amsterdam, North-Holland (2000).

\bibitem{Balestriero2019} Balestriero, R., Cosentino, R., Aazhang, B., and Baraniuk R.G.: The Geometry of Deep Networks: Power Diagram Subdivision. 33rd Conference on Neural Information Processing Systems (NeurIPS 2019), Vancouver, Canada, 1-10 (2019).

\bibitem{Barrena2005} Barrena, R., Ramella, M., Boschin, W. et al.: VGCF detection of galaxy systems at intermediate redshifts. Astron. Astrophys. \textbf{444(3)}, 685-695 (2005).

\bibitem{Borovsky2018} Borovsky, J.E.: The spatial structure of the oncoming solar wind at Earth and the shortcomings of a solar-wind monitor at L1. Journal of Atmospheric and Solar-Terrestrial Physics \textbf{177}, 2-11 (2018).

\bibitem{Blanton2005} Blanton, M.R., Eisenstein, D., Hogg, D.W., et al.: Relationship between Environment and the Broadband Optical Properties of Galaxies in the Sloan Digital Sky Survey. Astrophys. J. \textbf{629}, 143-157 (2005).

\bibitem{Burton2013} Burton, C.S., Jarvis, M.J., Smith, D.J.B. et al.: Herschel-ATLAS/GAMA: the environmental density of far-infrared bright galaxies at z {\ensuremath{\leq}} 0.5. Mon. Notic. R. Astron. Soc. \textbf{433(1)}, 771-786 (2013).

 \bibitem{Busch2020} Busch, P., White, S.D.M.: The Tessellation-Level-Tree: characterizing the nested hierarchy of density peaks and their spatial distribution in cosmological N-body simulations. Mon. Notic. R. Astron. Soc. \textbf{493}, 5693-5712 (2020).

\bibitem{Cabrera2008} Cabrera, G. F., Casassus, S., Hitschfeld, N. et al.: Bayesian Image Reconstruction Based on Voronoi Diagrams. Astrophys. J. \textbf{672}, 1272-1285 (2008).

\bibitem{Cappellari2002} Cappellari, M., Copin, Y.: Adaptive Spatial Binning of 2D Spectra and Images Using Voronoi Tessellations. Galaxies: The Third Dimension, ASP Conference Proceedings. \textbf{282}, p.515 (2002). 

\bibitem{Cappellari2003} Cappellari, M., Copin, Y.: Adaptive spatial binning of integral-field spectroscopic data using Voronoi tessellations. Mon. Notic. R. Astron. Soc. \textbf{342}, 345-354 (2003).

\bibitem{Chadha2016} Chadha, A., Andreopoulos, Y.: Voronoi-based compact image descriptors: Efficient Region-of-Interest retrieval with VLAD and deep-learning-based descriptors. eprint arXiv:1611.08906, (2016). 

\bibitem{Chiaki2016} Chiaki, G., Yoshida, N., Hirano, S.: Gravitational collapse and the thermal evolution of low-metallicity gas clouds in the early Universe. Mon. Notic. Roy. Astron. Soc. \textbf{463(3)}, 2781-2798 (2016).

\bibitem{Chilingarian2010} Chilingarian, I.V., Melchior, A.L., Zolotukhin, I.Yu.: Analytical approximations of K-corrections in optical and near-infrared bands. Mon. Notic. R. Astron. Soc. \textbf{405}, 1409-1420 (2010).

\bibitem{Coles1990} Coles, P., Barrow, J.D.: Microwave background constraines on the Voronoi model of large-scale structure. Mon. Notic. R. Astron. Soc. \textbf{244}, 557-562 (1990).

\bibitem{Cooper2005}  Cooper, M.C., Newman, J.A., Madgwick, D.S., et al.: Microwave background constraines on the Voronoi model of large-scale structure. Astrophys. J. \textbf{634(2)}, 833-848 (2005).

\bibitem{Coutinho2016} Coutinho, B.C., Hong, S., Albrect, K. et al.: The network behind the cosmic web. arXiv:1604.03236v2 (2016).

\bibitem{Cubul2014} Cybulski, R., Yun, Min S., Fazio, G.G. et al.: From voids to Coma: the prevalence of pre-processing in the local Universe. Mon. Notic. Roy. Astron. Soc. \textbf{439(4)}, 3564-3586 (2014).

\bibitem{Cucciati2006} Cucciati, O., Iovino, A., Marinoni, C., et al.: The VIMOS VLT Deep Survey: the build-up of the colour-density relation. Astron. Astrophys. \textbf{458}, 39-52 (2006).

\bibitem{Debnath2015} Debnath, D., Gainer, J.S., Kim, D., and Matchev, K.: Edge Detecting New Physics the Voronoi Way. arXiv:1506.04141 (2015).

\bibitem{Debnath2018} Debnath, D.: Generic and sensitive searches for new physic. A dissertation presented to the graduate school of the University of Florida. 293 p. (2018). 

\bibitem{Descartes1644} Descartes R.: Principia Philosophiae. Amsterdam, Ludovicus Elzevirius (1644).

\bibitem{Diehl2006} Diehl, S., Statler, T.S.: Adaptive binning of X-ray data with weighted Voronoi tessellations. Mon. Notic. R. Astron. Soc. \textbf{368}, 497–510 (2006). 

\bibitem{Dobrycheva2013}  Dobrycheva, D.V.: The New Galaxy Sample from SDSS DR9 at 0.003 $<$ z $<$ 0.1. Odessa Astron. Publ. \textbf{26}, 187-189 (2013).

\bibitem{Dobrycheva2015}  Dobrycheva, D.V., Melnyk, O.V., Vavilova, I.B., Elyiv, A.A.: Environmental Density vs. Colour Indices of the Low Redshifts Galaxies. Astrophysics \textbf{58}, 168-180 (2015).

\bibitem{Dobrycheva2016}  Dobrycheva, D.V., Vavilova, I.B.: No the Holmberg Effect for Galaxy Pairs Selected from the SDSS DR9 at z {$\leq$} 0.06. Odessa Astron. Publ. \textbf{29}, 37-41  (2016).

\bibitem{Dobrycheva2017b}  Dobrycheva, D.V.: Morphological content and color indices bimodality of a new galaxy sample at the redshifts z \&lt; 0.1. PhD Thesis in Phys.-Math. Sciences, Main Astronomical Observatory, NAS of Ukraine, 132pp. (2017).

\bibitem{Dobrycheva2017}  Dobrycheva, D.V., Vavilova, I.B., Melnyk, O.V., Elyiv, A.A.: Machine learning technique for morphological classification of galaxies at z<0.1 from the SDSS. arXiv.1712.08955 (2017). 
 
\bibitem{Doroshkevich1997} Doroshkevich, A., Gottlober, S., Madsen S.: The accuracy of parameters determined with the core-sampling method: applications to Voronoi tessellations. Astron. Astrophys. Suppl. Ser. \textbf{123}, 495-506 (1997). 

\bibitem{Dressler1980} Dressler, A.: Galaxy morphology in rich clusters -- Implications for the formation and evolution of galaxies. Astrophys. J. \textbf{236}, 351-365 (1980).
 
\bibitem{Ebeling1993}  Ebeling, H., Wiedenmann, G.: Detecting structure in two dimensions combining Voronoi tessellation and percolation. Physical Review E (Statistical Physics, Plasmas, Fluids, and Related Interdisciplinary Topics) \textbf{47(1)}, 704-710 (1993). 

\bibitem{Elyiv2009}  Elyiv, A., Melnyk, O., Vavilova, I.: High-order 3D Voronoi tessellation for identifying isolated galaxies, pairs and triplets. Mon. Notic. R. Astron. Soc. \textbf{394}, 1409-1418 (2009).

\bibitem{Elyiv2015} Elyiv, A., Marulli, F., Pollina, G. et al.: Cosmic voids detection without density measurements. Mon. Notic. R. Astron. Soc. \textbf{448}, 642-653 (2015).

\bibitem{Elyiv2020}  Elyiv, A.A., Melnyk, O.V., Vavilova, I.B. et al.: Machine-learning computation of distance modulus for local galaxies. Astron. Astrophys. \textbf{635}, id.A124, 7 pp. (2020).

\bibitem{Fu2012} Fu, Z. and Liu, S.: a Vomr-Tree Based Parallel Range Query Method on Distributed Spatial Database. ISPRS Annals of Photogrammetry, Remote Sensing and Spatial Information Sciences, \textbf{12}, 37-43 (2012).
 
\bibitem{Gerke2005}  Gerke, B.F., Newman, J.A., Davis, M. et al.:  The DEEP2 Galaxy Redshift Survey: First Results on Galaxy Groups. Astrophys. J. \textbf{625(1)}, 6-22 (2005). 

\bibitem{Gregul1991}  Gregul, A.Ia., Mandzhos, A.V., Vavilova, I.B.: The existence of the structural anisotropy of the Jagiellonian field of the galaxies. Astrophys. Space Sci.  \textbf{185}, 223-235 (1991).

\bibitem{Grokhovskaya2019} Grokhovskaya, A.A., Dodonov, S.N.: Large Scale Distribution of Galaxies in The Field HS 47.5-22. I. Data Analysis Technique. Astrophys. Bulletin.  \textbf{74}, 379-387 (2019).

\bibitem{Guedj2017} Guedj, B. and Srinivasa Desikan, B.: Pycobra: A Python Toolbox for Ensemble Learning and Visualisation. arXiv:1707.00558.

\bibitem{Hidding2015} Hidding, J., van de Weygaert, R., Vegter, G., and Jones, B.J.T.: Adhesion and the Geometry of the Cosmic Web. In: Thirteenth Marcel Grossmann Meeting: On Recent Developments in Theoretical and Experimental General Relativity, Astrophysics and Relativistic Field Theories, p. 2142-2144 (2015).

\bibitem{Hettiarachchi2016} Hettiarachchi, R. and Peters, J.F.: Voronoi Region-Based Adaptive Unsupervised Color Image Segmentation. arXiv:1604.00533 (2016).

\bibitem{Hogg2004} Hogg, D.W., Blanton, M.R., Brinkmann, J., et al: The Dependence on Environment of the Color-Magnitude Relation of Galaxies. Astrophys. J. \textbf{601}, L29 (2004).

\bibitem{Honda2019} Honda, C., Yasuda, Y., Yokota, Y.: Lunar Crater Spatial Distribution for Each Surface Model Age. American Geophysical Union, Fall Meeting 2019, abstract #P31C-3473 (2019).

\bibitem{Hubber2016} Hubber, D.A., Ercolano, B., Dale, J.: Observing gas and dust in simulations of star formation with Monte Carlo radiation transport on Voronoi meshes. Mon. Notic. Roy. Astron. Soc., \textbf{456(1)}, 756-766 (2016).

\bibitem{Hung2019} Hung, D., Lemaux, B.C., Gal, R.R., Tomczak, A.R. et al: Establishing a New Technique for Discovering Large-Scale Structure Using the ORELSE Survey. Mon. Notic. Roy. Astron. Soc., Advance Access, 1-39 (2019).

\bibitem{Icke1987} Icke, V., van de Weygaert, R.: Fragmenting the Universe. 1. Statistics of two-dimensional Voronoi foams. Astron. Astrophys. \textbf{184}, 16-32 (1987).

\bibitem{Icke1988} Icke, V., van de Weygaert, R.: Voronoi foam as a model of the medium-scale universe. Large-Scale Structures in the Universe Observational and Analytical Methods: Proceedings of a Workshop, Held at the Physikzentrum Bad Honnef. \textbf{310}, 257-266 (1988).

\bibitem{Icke1991} Icke, V., van de Weygaert, R.: The galaxy distribution as a Voronoi foam. Royal Astronomical Society, Quarterly Journal \textbf{32}, 85-112 (1991).

\bibitem{Ickeuchi1991} Ickeuchi, S., Turner, E.I.: Quasi-periodic structures  in the large-scale galaxy distribution and three-dimensional Voronoi tessellation. Mon. Notic. R. Astron. Soc. \textbf{250}, 519-522 (1991).

\bibitem{Igashov2020} Igashov, I., Olechnovic K., Kadukova, M. et al: VoroCNN: Deep convolutional neural network built on 3D Voronoi tessellation of protein structures. BioArxiv:063586v1 (2020).

\bibitem{Inkulu2011} Inkulu, R. and Kapoor, S.: ANN queries: covering Voronoi diagram with hyperboxes. arXiv:1111.5893.

\bibitem{Jackson1972} Jackson, J.C.: A critique of Rees's theory of primordial gravitational radiation.  Mon. Notic. Roy. Astron. Soc., \textbf{156}, 1P (1972).

\bibitem{Karachentsev1996} Karachentsev, I. D., Makarov, D.A.: The Galaxy Motion Relative to Nearby Galaxies and the Local Velocity Field. Astron. J. \textbf{111}, p. 794 (1996).

\bibitem{Kim2002} Kim, R.S.J., Kepner, J.V., Postman, M. et al.: Detecting Clusters of Galaxies in the Sloan Digital Sky Survey. I. Monte Carlo Comparison of Cluster Detection Algorithms. Astron. J. \textbf{123(1)}, 20-36 (2002). 

\bibitem{Lam2019} Lam, M.C., Hambly, N.C., Rowell, N., Chambers, K.C. et al.: The white dwarf luminosity functions from the Pan-STARRS 1 3{\ensuremath{\pi}} Steradian Survey. Mon. Notic. R. Astron. Soc., \textbf{482(1)}, 715-731 (2019). 

\bibitem{Lang2015} Lang, M., Holley-Bockelmann, K., Sinha, M. et al.: Voronoi Tessellation and Non-parametric Halo Concentration. Astrophys. J., \textbf{811(2)}, 9 pp. (2015). 

\bibitem{Lindenbergh2002} Lindenbergh, R.: Limits of Voronoi Diagrams. PhD thesis, 132 (2002). 

\bibitem{Ling1987} Ling, E.N.: New Statistical Approaches to Galaxy Clustering. PhD thesis, Sussex Univ., Brighton (England) (1987).

\bibitem{Lopes2004}  Lopes, P.A.A., de Carvalho, R.R., Gal, R.R. et al.:The Northern Sky Optical Cluster Survey. IV. An Intermediate-Redshift Galaxy Cluster Catalog and the Comparison of Two Detection Algorithms. Astron. J. \textbf{128(3)}, 1017-1045 (2004). 

\bibitem{Marinoni2002}  Marinoni, C., Davis, M., Newman, J.A. et al.: Three-dimensional Identification and Reconstruction of Galaxy Systems within Flux-limited Redshift Surveys. Astrophys. J. \textbf{580(1)}, 122-143 (2002).

\bibitem{Matsuda1984} Matsuda, T., Shima, E.: Topology of Supercluster-Void Structure. Progress of Theoretical Physics \textbf{71}, 855-858 (1984). 

\bibitem{Melnyk2006b}  Melnyk, O.V., Elyiv, A.A., Vavilova, I.B.: The structure of the Local Supercluster of galaxies detected by three-dimensional Voronoi's tessellation method. Kinemat. Fiz. Neb. Tel \textbf{22}, 283-296 (2006).   
   
\bibitem{Melnyk2006c}  Melnyk, O.V., Elyiv, A.A., Vavilova, I.B.: 3-D Voronoi's Tessellation as a Tool for Identifying Galaxy Groups. Galaxy Evolution Across the Hubble Time, Edited by F. Combes and J. Palous, Proceedings of IAU Symposium №235, pp.223-223 (2006). 
   
\bibitem{Melnyk2009}  Melnyk, O.V., Elyiv, A.A., Vavilova, I.B.: Mass-to-light ratios for galaxy pairs and triplets in various environments. Kinemat. Phys. Celest. Bodies. \textbf{25}, 43-47 (2009). 
   
\bibitem{Melnyk2012}  Melnyk, O.V., Dobrycheva, D.V., Vavilova, I.B.: Morphology and color indices of galaxies in Pairs: Criteria for the classification of galaxies. Astrophysics \textbf{55}, 293--305 (2012).

\bibitem{Neyrinck2008} Neyrinck, M.C.: ZOBOV: a parameter-free void-finding algorithm. Mon. Notic. R. Astron. Soc. \textbf{386}, 2101-2109 (2008). 

\bibitem{Neyrinck2008b} Neyrinck, M.C.: The Cosmic Spiderweb and General Origami Tessellation Design. arXiv:1809.00015 (2008).

\bibitem{OMill2012} O'Mill, A.L., Duplancic, F., Lambas, G.D. et al.: Galaxy triplets in Sloan Digital Sky Survey Data Release 7 - I. Catalogue. Mon. Notic. R. Astron. Soc. \textbf{421}, 1897-1907 (2012). 

\bibitem{Panko2006} Panko, E., Flin, P.: A Catalogue of Galaxy Clusters and Groups Based on the Muenster Red Sky Survey. The Journal of Astronomical Data. \textbf{12}, 1P (2006). 

\bibitem{Paranjape2020} Paranjape, A. and Alam, S.: Voronoi volume function: a new probe of cosmology and galaxy evolution. Mon. Notic. R. Astron. Soc. \textbf{495}, 3233-3251 (2020).

\bibitem{Park2007} Park, C., Choi, Y.-Y., Vogeley, M.S. et al.: Environmental Dependence of Properties of Galaxies in the Sloan Digital Sky Surveyy. Astrophys. J. \textbf{658}, 898-916 (2007).

\bibitem{Pereira2018) Pereira, M.E.S., Soares-Santos, M., Makler, M. et al. Weak-lensing calibration of a stellar mass-based mass proxy for redMaPPer and Voronoi Tessellation clusters in SDSS Stripe 82. arXiv:1708.03329, 1361-1372 (2018).

\bibitem{Pereira2017a} Pereira, S., Campusano, L.E., Hitschfeld-Kahler, N. et al. A 3D Voronoi+Gapper Galaxy Cluster Finder in Redshift Space to z{\ensuremath{\approx}} 0.2 I: An Algorithm Optimized for the 2dFGRS}. Astrophys. J., \textbf{838}, 2, 109 (2017).
     
\bibitem{Pereira2017b} Pereira, S., Campusano, L., Hitschfeld-Kahler, N.et al.: A 3D Voronoi + Gapper Galaxy Cluster Finder in Redshift Space to z{\ensuremath{\approx}} 0.2 I: An Algorithm Optimized for the 2dFGRS. Astrophys. J. \textbf{838(2)}, 18 pp. (2017).

\bibitem{Pratsyovity2018} Pratsyovity, M.V., Syta, H.M.: Geometric mosaics of the Great Ukrainian (to the 150th anniversary of Professor G. Voronoi). Visnyk of the NAS of Ukraine. \textbf{4}, 92-101 (2018). 

\bibitem{Pulatova2015}  Pulatova, N.G., Vavilova, I.B., Sawangwit, U. et al.: The 2MIG isolated AGNs - I. General and multiwavelength properties of AGNs and host galaxies in the northern sky. Mon. Notic. R. Astron. Soc. \textbf{447}, 2209-2223 (2015). 

\bibitem{Ramella2001} Ramella, M., Boschin, W., Fadda, D., Nonino, M.: Finding galaxy cluster using Voronoi tessellations. Astron. Astrophys. \textbf{368}, 776-786 (2001).

\bibitem{SanRoman2018} San Roman, I., Cenarro, A.J., D{\'\i}az-Garc{\'\i}a, L.A. et al.: The ALHAMBRA survey: 2D analysis of the stellar populations in massive early-type galaxies at z < 0.3. Astron. Astrophys. \textbf{609}, A20.

\bibitem{SantiagoBautista2020} Santiago-Bautista, I., Caretta, C.A., Bravo-Alfaro, H. et al.: Identification of filamentary structures in the environment of superclusters of galaxies in the Local Universe. Astron. Astrophys. \textbf{637}, id.A31, 26 pp. (2020).

\bibitem{Schlegel1998} Schlegel, D.J., Finkbeiner, D.P., Davis, M.: Maps of Dust Infrared Emission for Use in Estimation of Reddening and Cosmic Microwave Background Radiation Foregrounds. Astrophys. J. \textbf{500}, 525-553 (1998).

\bibitem{Scoville2013} Scoville, N., Arnouts, S., Aussel, H. et al.: Evolution of Galaxies and Their Environments at z = 0.1-3 in COSMOS. Astrophys. J. Suppl. \textbf{206(1)}, 3 (2013).

\bibitem{Sheth2004} Sheth, R.K., van de Weygaert, R.: A hierarchy of voids: much ado about nothing. Mon. Notic. R. Astron. Soc.  \textbf{350}, 517-538 (2004). 

\bibitem{Soares2011} Soares-Santos, M., de Carvalho, R.R., Annis, J., Gal, R.R. et al.: The Voronoi Tessellation Cluster Finder in 2+1 Dimensions. Astrophys. J. \textbf{727(1)}, 14 pp. (2011).

\bibitem{Sochting2012} S{\"o}chting, I.K., Coldwell, G.V., Clowes, R.G. et al.: Ultra Deep Catalogue of Galaxy Structures in the Cosmic Evolution Survey field. Mon. Notic. R. Astron. Soc. \textbf{423}, 2436-2450 (2012).

\bibitem{Springel2010} Springel, V.: E pur si muove: Galilean-invariant cosmological hydrodynamical simulations on a moving mesh. Mon. Notic. R. Astron. Soc. \textbf{401}, 791-851 (2010). 

\bibitem{Subba1992}  Subba, R.M.U., Szalay, A.S.: Statistics of pencil-beams in Voronoi foams. Astrophys. J. \textbf{391}, 483-493 (1992). 

\bibitem{Sutter2015} Sutter, P.M., Lavaux, G., Hamaus, N., Pisani, A. et al.: VIDE: The Void IDentification and Examination toolkit. Astronomy and Computing \textbf{9}, 1-9 (2015).

\bibitem{Syta2009} Syta, H., van de Weygaert, R.: Life and Times of Georgy Voronoi. arXiv:0912.3269 (2009).

\bibitem{Tal2014} Tal, T., Dekel, A., Oesch, P., et al.: Observations of Environmental Quenching in Groups in the 11 GYR since $z = 2.5$: Different Quenching for Central and Satellite Galaxies. Astrophys. J. \textbf{789}, 1-11 (2014).

\bibitem{Trimble2014} Trimble, V.: Nor yet the last to lay the old aside: Structuring the Something. In: American Astronomical Society Meeting Abstracts, \textbf{223}, 90.01 (2014).

\bibitem{Vavilova1995}  Vavilova, I.B.: An investigation of large-scale galaxy distribution in the Local Supercluster and the Jagellonian Field by the methods of cluster, fractal and wavelet analysis. PhD Thesis in Phys.-Math. Sciences, Main Astronomical Observatory, NAS of Ukraine, 222 p. (1995). 

\bibitem{Vavilova2005b}  Vavilova, I., Melnyk, O.: Voronoi tessellation for galaxy distribution. Proceedings of the Third Voronoi Conference on Analytic Number Theory and Spatial Tessellations. \textbf{55}, 203-212 (2005). 

\bibitem{Vavilova2005c}  Vavilova, I.B., Karachentseva, V.E., Makarov, D.I., Melnyk, O.V.: Triplets of Galaxies in the Local Supercluster. I. Kinematic and Virial Parameters. Kinemat. Fiz. Neb. Tel. \textbf{1}, 3-20 (2005). 
   
\bibitem{Vavilova2009}  Vavilova, I.B., Melnyk, O.V., Elyiv, A.A.: Morphological properties of isolated galaxies vs. isolation criteria. Astron. Nachr. \textbf{330}, 1004-1009 (2009). 

\bibitem{Vavilova2015}  Vavilova, I.B., Bolotin, Yu.L., Boyarsky, A.M. et al.: Dark matter: Observational manifestation and experimental searches. Akademperiodyka, Kyiv (2015).

\bibitem{Vavilova2018}  Vavilova, I.B., Elyiv, A.A., Vasylenko, M.Yu.: Behind the Zone of Avoidance of the Milky Way: what can we Restore by Direct and Indirect Methods? Radiophysics and Radioastronomy \textbf{23}, 244-257 (2018). 

\bibitem{Vavilova2020a} Vavilova, I., Pakuliak, L., Babyk, I. et al.: Surveys, Catalogues, Databases, and Archives of Astronomical Data. In: Knowledge Discovery in Big Data from Astronomy and Earth Observation, Eds. P. Scoda and F. Adam, Elsevier. pp. 57-102 (2020).

\bibitem{Vavilova2020b} Vavilova, I., Dobrycheva, D., Vasylenko, M. et al.: Multiwavelength Extragalactic Surveys: Examples of Data Mining. Ln: Knowledge Discovery in Big Data from Astronomy and Earth Observation,  Eds. P. Scoda and F. Adam, Elsevier. pp. 307-323 (2020).

\bibitem{Vavilova2020c}  Vavilova, I.B.: Astroinformatics of the Large-Scale Structures of the Universe. Dr. Hab. Thesis in Phys.-Math. Sciences, Main Astronomical Observatory, NAS of Ukraine, 388 p. (2020).

\bibitem{Voronoi1907} Voronoi, G.: Nouvelles applications des parameterscontinus a la theorie des formes quadratques. Premier Memorie.  Sur quelques proprietes  des formes quadratiques positive parfaites.  J. reine angew. Math. \textbf{133(2)} 97-156 (1907); \textbf{133(3)}, 157-158 (1907). 

\bibitem{Voronoi1908} Voronoi, G.: Nouvelles applications des parameterscontinus a la theorie des formes quadratques. Deuxieme Memorie. Recherches sur les paralleloedres primitives. J. reine angew. Math. \textbf{134(3)} 198-246 (1908); \textbf{134(4)}, 247-287 (1908); \textbf{136(2)}, 67-178 (1909). 

\bibitem{Weinberger2020} Weinberger, R., Springel, V., Pakmor, R.: The AREPO Public Code Release. Astrophys. J. Suppl., \textbf{248(2)}, 32 (2020).

\bibitem{Weygaert1989} van de Weygaert, R., Icke, V.: Fragmenting the universe. II - Voronoi vertices as Abell clusters. Astron. Astrophys. \textbf{213}, 1-9 (1989).

\bibitem{Weygaert2009} van de Weygaert, R., Aragon-Calvo, M.A., Jones, B.J.T. et al.: Geometry and Morphology of the Cosmic Web: Analyzing Spatial Patterns in the Universe. arXiv:0912.3448 (2009).

\bibitem{Ying2015} Ying, S., Guang, X., Chengpeng, L., et al.: Point Cluster Analysis Using a 3D Voronoi Diagram with Applications in Point Cloud Segmentation. ISPRS International Journal of Geo-Information. \textbf{4}, 1480-1499 (2015). 

\bibitem{Yoshioka1989} Yoshioka, S. and Ikeuchi, S.: The Large-Scale Structure of the Universe and the Division of Space. Astrophys. J., \textbf{341}, 16-25 (1989).
      
\bibitem{Zaninetti1989} Zaninetti, L.: Dynamical Voronoi tessellation. I. Astron. Astrophys. \textbf{224}, 345-350 (1989). 

\bibitem{Zaninetti1990} Zaninetti, L.: Dynamical Voronoi tessellation. II. Astron. Astrophys. \textbf{233}, 293-300 (1990). 

\bibitem{Zaninetti2010a} Zaninetti, L.: Practical Statistics for the Voids Between Galaxies. Serbian Astronomical Journal. \textbf{181}, 19-29 (2010). 

\bibitem{Zaninetti2010b} Zaninetti, L.: A geometrical model for the catalogs of galaxies. Revista Mexicana de Astronomía y Astrofísica \textbf{46}, 115-134 (2010).     
       
\bibitem{Zaninetti2012} Zaninetti, L.: New Analytic Results for Poissonian and non-Poissonian Statistics of Cosmic Voids. Revista Mexicana de Astronomía y Astrofísica. \textbf{48}, 209-222 (2012).  
    
\bibitem{Zaninetti2013} Zaninetti, L.: Photometric Effects and Voronoi-Diagrams as a Mixed Model for the Spatial Distribution of Galaxies. The Open Astronomy Journal. \textbf{6}, 48-71 (2013). 

\end{thebibliography}
\end{document}